\documentclass[fleqn,usenatbib]{mnras}

\usepackage{newtxtext,newtxmath}

\usepackage[T1]{fontenc}
\usepackage{ae,aecompl}
\usepackage{ulem}

\usepackage{graphicx}	
\usepackage{amsmath}	
\usepackage{comment}
\usepackage{soul}
\usepackage{xcolor}

\newcommand{\cmfast}{{\tt 21cmFAST}} 
\newcommand{\cmmc}{{\tt 21cmMC}}

\newcommand\lsim{\mathrel{\rlap{\lower4pt\hbox{\hskip1pt$\sim$}}
        \raise1pt\hbox{$<$}}}
\newcommand\gsim{\mathrel{\rlap{\lower4pt\hbox{\hskip1pt$\sim$}}
        \raise1pt\hbox{$>$}}}
\def\myputfigure#1#2#3#4#5%
{\vskip#5pt\makebox[0pt]{\hskip#2in
\includegraphics[width=#3\textwidth]{#1}}\vskip#4pt\hfill}

\title[On-the-fly ionizing photon non-conservation correction]{Calibrating excursion set reionization models to approximately conserve ionizing photons}

\author[J. Park et al.]{
Jaehong~Park,$^{1}$\thanks{E-mail: jaehongpark@kias.re.kr (JP)}
Bradley Greig,$^{2,3}$
Andrei Mesinger,$^{4}$
\\
$^{1}$School of Physics, Korea Institute for Advanced Study (KIAS), 85 Hoegiro, Dongdaemun-gu, Seoul 02455, Republic of Korea\\
$^{2}$ARC Centre of Excellence for All-Sky Astrophysics in 3 Dimensions (ASTRO 3D), University of Melbourne, VIC 3010, Australia\\
$^{3}$School of Physics, The University of Melbourne, Parkville, VIC 3010, Australia\\
$^{4}$Scuola Normale Superiore, Piazza dei Cavalieri 7, I-56126 Pisa, Italy\\
}

\date{Accepted XXX. Received YYY; in original form ZZZ}

\pubyear{2021}

\hypersetup{draft}
\begin{document}
\label{firstpage}
\pagerange{\pageref{firstpage}--\pageref{lastpage}}
\maketitle

\begin{abstract}
The excursion set reionization framework is widely used, due to its speed and accuracy in reproducing the 3D topology of reionization.  However, it is known that it does not conserve photon number. Here, we introduce an efficient, on-the-fly recipe to approximately account for photon conservation. Using a flexible galaxy model shown to reproduce current high-$z$ observables, we quantify the bias in the inferred reionization history and galaxy properties resulting from the non-conservation of ionizing photons. Using a mock 21-cm observation, we perform inference with and without correcting for ionizing photon conservation. We find that ignoring photon conservation results in very modest biases in the inferred galaxy properties, for our fiducial model. The notable exception is in the power-law scaling of the ionizing escape fraction with halo mass, which can be biased from the true value by $\sim 2.4\sigma$ (corresponding to $\sim -0.2$ in the power-law index). Our scheme is implemented in the public code {\cmfast}.
 \end{abstract}




%
%
\section{Introduction}\label{sec:intro}

Due to the short mean free paths of Lyman limit photons in the neutral intergalactic medium (IGM), cosmic reionization is inherently inhomogeneous. Ionized regions surrounding the first galaxies expand into the neutral IGM, growing and percolating as new galaxies are formed \citep[e.g.][]{Friedrich2011,Furlanetto&Oh2016}. This bimodality of the ionization state of the IGM motivated the development of excursion set models of reionization \citep{Furlanetto2004,Paranjape2016}. In the excursion set framework, the cumulative number of ionizing photons is compared to the number of baryons plus recombinations, in spheres of decreasing radii around a given patch of the IGM. If there are more ionizing photons, the patch is determined to be ionized.

These models can be easily implemented on 3D grids using a series of fast Fourier transforms (FFTs), allowing for fast realizations of the reionization topology on large scales. This has made excursion set models very popular in semi-numerical Epoch of Reionization (EoR) simulations, allowing us to efficiently explore the large-parameter space of astrophysical and cosmological uncertainties.  Excursion set models have been tested numerous times against more accurate radiative transfer (RT) simulations (\citealt{Zahn2007, Zahn2011,21cmfast,Majumdar2014,Hutter2018, Thelie2021}). When compared at a fixed stage of the EoR (i.e. a given filling factor of $\ion{H}{II}$ regions, $Q_{\rm \ion{H}{II}}$), excursion set and more accurate RT approaches result in ionization maps that agree well on moderate to large scales ($\gtrsim$ Mpc).

However, it is well known that excursion set approaches do not conserve the ionizing photon number \citep{Zahn2007,Zahn2011, Paranjape2016, Choudhury2018,Hutter2018}. Fundamentally, this problem occurs when neighboring $\ion{H}{II}$ regions overlap. Photons in excess of the criteria for ionization are not correctly re-distributed into neighboring cells \citep{Zahn2007}. Therefore, without accounting for this non-conservation when using semi-numerical EoR simulations to infer galaxy properties from observations, we expect a bias in the recovered ionizing emissivity (e.g. an overestimate of the ionizing photon escape fraction; \citealt{Hutter2018, Molaro2019}).

Several approaches have been developed to conserve ionizing photons in approximate radiative transfer schemes \citep[e.g.][]{Paranjape2016,Kim2016,Hassan2017,Choudhury2018,Molaro2019}. However, these can come at the cost of significantly increased computation time and/or stepping away from the well-tested excursion set framework entirely.

Motivated by the need for rapid parameter space exploration, here we develop a computationally efficient, approximate correction for ionizing photon conservation in 3D excursion set models. Rather than explicitly tracking the number of ionizing photons, we calibrate the simulated EoR history to match an analytically derived EoR history. We effectively re-scale (increase) the ionizing emissivity as a function of redshift to compensate for the loss of ionizing photons. This calibration therefore preserves the well-tested reionization topology, at a fixed stage of the EoR, resulting from excursion set models \citep{Zahn2007,Zahn2011,21cmfast,Hutter2018, Thelie2021}.
 
We also quantify the biases in the recovered astrophysical parameters when not accounting for the photon conservation using a mock observation of the 21-cm signal. We infer the astrophysical parameters within a fully Bayesian framework using the public MCMC sampler of 3D EoR simulations {\cmmc}\footnote{https://github.com/21cmfast/21CMMC} \citep{21CMMC,Greig2017,Greig2018}. Our correction has been implemented in the  (v3.0\footnote{https://github.com/21cmfast/21cmFAST}; \citealp{21cmfast_v3}) release of {\cmfast} \citep{Mesinger2007,21cmfast}.

This paper is organized as follows. In \S~\ref{sec:ionization_field} we summarize how {\cmfast} computes ionization fields before we introduce the methodology for correcting the photon non-conservation in \S~\ref{sec:correction}. In \S~\ref{sec:mock_obs}, we outline the set-up for Bayesian inference of astrophysical parameters, and quantify biases from photon non-conservation in \S~\ref{sec:results}. Finally, we summarize our results in \S~\ref{sec:conclusion}. We assume a standard ${\rm \Lambda}$CDM cosmology based on {\it Planck} 2016  \citep{PlanckXIII}: ($h$, $\Omega_{\rm m}$, $\Omega_{\rm b}$, $\Omega_{\Lambda}$, $\sigma_{8}$, $n_{\rm s}$)=(0.678, 0.308, 0.0484, 0.692, 0.815, 0.968).  Unless stated otherwise, we quote all quantities in comoving units, and  UV magnitudes refer to the rest-frame 1500{\AA}  AB magnitude.

%
%
\section{Ionization fields from the excursion set}\label{sec:ionization_field}

Here we briefly summarize the excursion set EoR model. The general analytic framework was first developed by  \cite{Furlanetto2004}, and subsequently applied to 3D realizations in \citet{Zahn2005, Mesinger2007}. Since then various semi-numerical simulations have adopted the excursion set framework when generating reionization fields \citep[e.g.][]{Zahn2005, Mesinger2007, Geil&Wyithe2008, Alvarez2009, Choudhury2009, Santos2010, 21cmfast, Visbal12, Mutch2016, Hutter2018}. Below we focus our discussion on its implementation in the public {\cmfast} code.

As mentioned above, the excursion set model divides the IGM during the EoR into (mostly) ionized and (mostly) neutral regions\footnote{The neutral IGM component could be partially ionized by X-ray photons that have long mean free paths. These partial ionizations are followed for each cell by integrating the emissivity back along the lightcone \citep{21cmfast}. Furthermore, the ionized IGM component has some residual neutral gas that depends on photo-ionization equilibrium with the local UV background (UVB), accounting also for self-shielding \citep{SM14}.}.
To determine whether or not an IGM patch is ionized, the cumulative number of ionizing photons is compared to the number of atoms plus recombinations.
Specifically, a voxel at spatial location and redshift $({\bf x}, z)$ is flagged as ionized if the following criterion is satisfied: 
\begin{equation}\label{eq:ionized}
n_{\rm ion}(\mathbf{x},z| R, \delta_{\rm R}) \geq (1 + n_{\rm rec})(1 - x_{\rm e}).
\end{equation}
Here $n_{\rm rec}$ is the cumulative number of recombinations per baryon, $x_{\rm e}$ is the fraction of ionizations by X-rays. The criterion in equation (\ref{eq:ionized}) is checked in spheres of decreasing radii starting from the maximum radius (set by the mean free path through the ionized IGM) to a radius equivalent in size to the cell resolution.  In {\cmfast}, the cumulative number of ionizing photons per baryon inside a spherical region of scale $R$ with overdensity $\delta_{\rm R}$ is given by
\begin{equation}\label{eq:nion}
n_{\rm ion} = \bar{\rho}^{-1}_{\rm b} \int_0^\infty {\rm d}M_{\rm h} \frac{{\rm d}n(M_{\rm h}, z | R, \delta_{\rm R})}{{\rm d}M_{\rm h}} f_{\rm duty} M_{\ast} f_{\rm esc} N_{\rm \gamma/b},
\end{equation}
where $\bar{\rho}_{\rm b}$ is the mean baryon density, ${\rm d}n(M_{\rm h},z|R,\delta_{\rm R})/{\rm d}M_{\rm h}$ is the conditional Press-Schechter halo mass function (HMF) 
normalized to the mean of the Sheth-Tormen HMF \citep{S-T1999} and $N_{\rm \gamma/b}$ is the number of ionizing photons per stellar baryon\footnote{We set $N_{\rm \gamma/b}=5000$, motivated by a Salpeter initial mass function \citep{Salpeter1955}. Note that this is degenerate with $f_{\rm esc}(M_h)$, which is treated as a free parameter when performing inference.}. $M_{\ast}$, $f_{\rm esc}$, $f_{\rm duty}$ are the stellar mass of a galaxy, the ionizing photon escape fraction, and the duty cycle, respectively, which will be described in \S~\ref{sec:galaxy_model}.

\begin{figure*}
    \includegraphics[width=17.5cm]{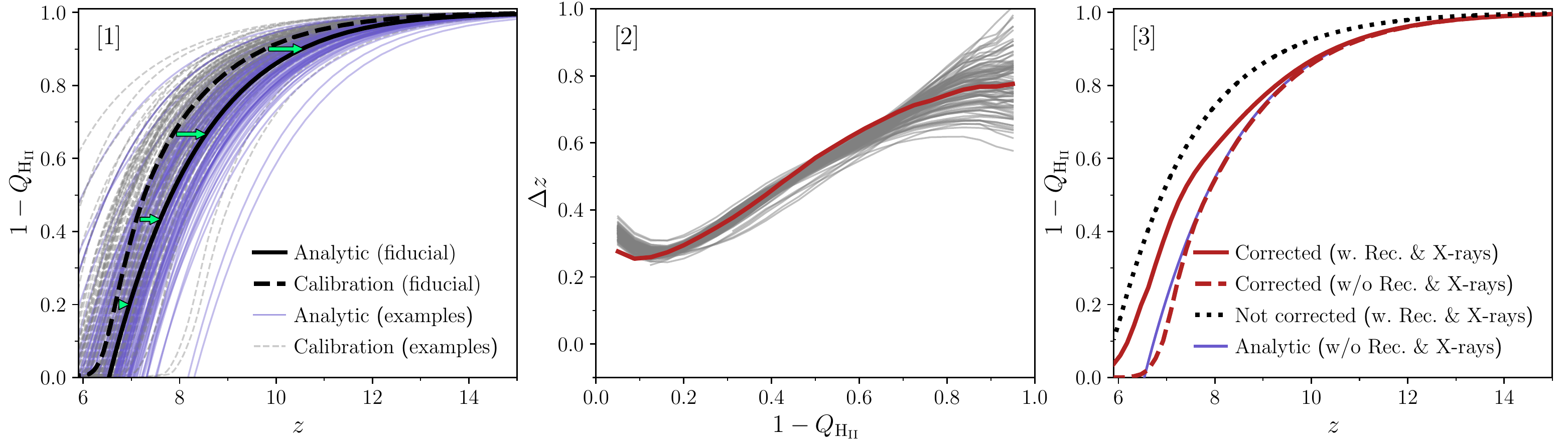}
    \caption{
    Each panel corresponds to a step of our calibration procedure described in the text. [1]: the EoR history from the analytic expression and the calibration curve (from {\cmfast}); both include only ionizations by UV sources. Thick lines correspond to the fiducial astrophysical parameters (listed in Table~\ref{Table:recovered_parameters}). Thin solid (violet) and thin dashed (grey) lines represent analytic EoR histories and calibration curves using 100 different combinations of parameters, which are randomly selected from the MCMC chain (see Sec.~\ref{sec:results}). Arrows (lime green) illustrate the calibration vector for the fiducial parameter set. [2]: the calibration vector, $\Delta z \equiv z_{\rm analytic}-z_{\rm calibration}$ (i.e. $\Delta z$ corresponds to the size of arrows in panel [1]). Thick solid (red) and thin (grey) lines represent $\Delta z$ for the fiducial parameter set and the 100 different parameter combinations, respectively. [3]: final EoR histories. The thick solid (red) line represent the corrected EoR history when considering inhomogeneous recombinations and X-rays. The thick dotted line shows the same EoR history, but without photon conservation (i.e. not corrected). We also show the corrected EoR histories when not considering X-rays and inhomogeneous recombinations (thick red dashed line), which is almost identical with the analytic EoR history (thin violet solid line).
    }
\label{fig:schematic}
\end{figure*}

{\cmfast} calculates inhomogeneous recombinations according to the sub-grid prescription of \citet{SM14}. The hydrogen recombination rate at a spatial position $\bf x$ and redshift $z$ is:
\begin{equation}\label{eq:Dn_recDt}
\frac{{\rm d}n_{\rm rec}}{{\rm d}t}(\mathbf{x},z) = \bar{n}_{\rm H} \alpha_{\rm B} \Delta_{\rm cell}^{-1} \int_0^{180} \left[ 1-x_{\ion{H}{I}}\right]^2 P_{\rm V} \Delta^2 {\rm d}\Delta\,,
\end{equation}
where $\Delta_{\rm cell} \equiv 1+\delta_{\rm nl}$ is density contrast of the cell, $P_{\rm V}$ is the volume-averaged probability distribution function of subgrid overdensity, $\Delta$, taken from \cite{Miralda-Escude2000} and adjusted for the mean contrast $\Delta_{\rm cell}$, $\alpha_{\rm B}$ is the case-B recombination coefficient and $x_{\ion{H}{I}}$ is the neutral fraction at the overdiensity $\Delta$ with the attenuation of the local, inhomogeneous ionizing background, $\Gamma$, computed using the fitting formula from \cite{Rahmati2013}. For further details, interested readers are referred to \citealp{SM14} (see also \citealp{Hutter2018}). 

%
%

\section{Ionizing photon non-conservation correction}\label{sec:correction}

The excursion set approach, outlined in equation~(\ref{eq:ionized}) does not explicitly follow the propagation of ionizing photons. Because the diffusion of ionizing photons when \ion{H}{II} regions overlap is not properly accounted for, the ionizing photon number is not conserved. This effect has been studied in detail, and depends on the specific implementation of the excursion set algorithm (e.g. \citealt{Zahn2007,Zahn2011, Paranjape2016, Choudhury2018,Hutter2018}). In the most common implementation described in the previous section, ionizing photons are effectively "lost", resulting in a slower EoR history. This might not be a problem if the ionizing efficiency or escape fraction are treated as effective (i.e. "tuning knob") parameters. However, physically interpreting the constraints on these parameters from observations requires understanding and correcting for the biases caused by ionizing photon non-conservation.

Here, we introduce an efficient method for calibrating excursion set algorithms on-the-fly using an exact analytic solution for a related EoR history. We estimate the (model-dependent) redshift delay resulting from the photon loss, shifting the excursion set EoR history according to this calibration curve. Our approach is motivated by the fact that the EoR topology from the excursion set has been extensively tested and shown to reproduce RT simulations, {\it at a given $Q_{\rm \ion{H}{II}}$} (e.g. \citealt{Zahn2007,Zahn2011,21cmfast,Hutter2018}). Our approach therefore preserves the EoR topology, shifting only the redshift associated with a given $Q_{\rm \ion{H}{II}}$. Below we describe the algorithm in more detail.

For a given choice of galaxy parameters (c.f. equation \ref{eq:nion}), we first simulate the reionization history using {\cmfast}
including only ionizations by UV sources (i.e. we do not consider recombinations, sinks of ionizing photons referred to as Lyman limit systems, or X-ray preionization at this stage). This EoR history is designated as the calibration curve. We then solve the corresponding evolution of the $\ion{H}{II}$ filling factor analytically:
\begin{equation}\label{eq:HII_filling_factor}
\frac{{\rm d}{Q_{\rm \ion{H}{II}}}}{{\rm d}t} = \frac{{\rm d}n_{\rm ion}}{{\rm d}t}
\end{equation}
We assume this analytic expression is the {\it true} EoR history, given our assumptions\footnote{Unfortunately, there is no straightforward analytic way of including inhomogeneous recombinations in the exact analytic solution as we cannot a-priory determine effective recombination times or sub-grid clumping factors. Commonly-used approximations such as constant clumping factors (e.g. \citealt{Maity2021}) do not capture the interplay of sources and sinks in our sub-grid recombination model (c.f. Figure 11 in \citealt{SM14}). Since we have to compare the excursion set and the exact analytic solution using the same assumptions, and iterating on the solution would be prohibitively expensive, we ignore recombinations in this first step.}.
In Figure~\ref{fig:schematic} (panel [1]), we show the analytic EoR history (thick black solid) and the calibration curve from {\cmfast} (thick black dashed) for our fiducial parameter set listed in Table~\ref{Table:recovered_parameters}. We also show the analytic EoR histories and calibration curves (thin grey and violet curves, respectively) for 100 samples of different astrophysical parameters corresponding to the posterior discussed in section \ref{sec:results}.  

Second, after obtaining the two EoR histories, we calculate the corresponding redshift calibration vector, $\Delta z(Q_{\rm \ion{H}{II}}) \equiv z_{\rm analytic}(Q_{\rm \ion{H}{II}})-z_{\rm calibration}(Q_{\rm \ion{H}{II}})$,
where $z_{\rm analytic}$ and $z_{\rm calibration}$ represent the redshift of the analytic EoR history from equation~(\ref{eq:HII_filling_factor}) and that of the calibration curve from {\cmfast}, respectively\footnote{To minimize numerical effects associated with the end stages of the EoR, where recombinations can be important, we linearly extrapolate the calibration vector $\Delta z$ above $Q_{\rm \ion{H}{II}} > 0.7$. We also note that the photon non-conservation correction is only applied once individual HII regions start overlapping \citep[e.g.][]{Friedrich2011,Furlanetto&Oh2016}, specifically at $Q_{\rm \ion{H}{II}} > 0.05$.}.
We show these calibration vectors in panel [2] of Fig.~\ref{fig:schematic} for the fiducial parameter sets (listed in Table~\ref{Table:recovered_parameters}, thick red curve) together with the 100 posterior samples (thin grey curves) as in [1]. We see $\Delta z$ is always positive, confirming that our implementation of the excursion set loses ionizing photons (i.e. has a delayed EoR history relative to the exact analytic solution; c.f. \citealt{Paranjape2014}). From the figure, we see that even without accounting for photon conservation, the impact on the EoR history is relatively modest: $\Delta z$ ranges approximately between 0.2 and 1.0. However, even such a relatively modest error can bias our interpretation of the ionizing escape fraction, as we quantify in Sect. \ref{sec:results}. It is also important to note that, despite the wide range of EoR histories shown in panel [1], the calibration curves all cluster together in a narrow region in this $\Delta z$ -- $Q_{\rm \ion{H}{II}}$ plane. This shows that the photon non-conservation is a self-similar process when defined as $\Delta z(Q_{\rm \ion{H}{II}})$. The fact that the calibration is mostly dependent on $Q_{\rm \ion{H}{II}}$ and only weakly depends on the details of the astrophysics is strongly suggestive that it can be applied more generally to models that include inhomogeneous recombinations for which we do not have an exact analytic solution (as we do in the next step).

Finally, we run the full {\cmfast} simulation, i.e. turning on all the desired physics such as inhomogeneous recombinations and X-ray ionizations. At any redshift, $z$, {\cmfast} computes all the relevant fields (e.g. density, velocity, spin temperature, recombination, etc.). We then apply the redshift calibration vector (c.f. panel [2]) to adjust the excursion set ionization field.  Specifically, we calculate the ionization field at an adjusted redshift, $z_{{\rm adj}} = z - \Delta z$, linearly evolving the density field to $z_{\rm adj}$ using the growth factor $D(z)$, i.e. $\delta({\rm x}, z_{\rm adj}) = \delta({\rm x},z) \times D(z_{\rm adj})/D(z)$. Effectively, this correction amounts to boosting the number of ionizing photons produced by increasing the effective number density of sources. The corrected EoR history is shown in [3] (thick red solid line) of Fig.~\ref{fig:schematic}, together with the EoR history without the correction (thick dotted line) and that of analytic EoR history (thin violet solid line) 

As we expect, the EoR history with photon conservation (thick red solid line in [3] of Fig.~\ref{fig:schematic}) is consistent with that of analytic EoR history at the early stage of reionization, because there are fewer overlapping regions at this stage. As reionization progresses, the corrected EoR history (i.e. with photon conservation) from {\cmfast} shows an offset from that of the analytic EoR history, because the former includes recombinations.\footnote{We note that there is a mild "knee" in the EoR history with photon conservation, due to the aforementioned extrapolation of the calibration curve at the late stages of reionization. As can be seen in [1] of Fig.~\ref{fig:schematic}, the calibration curve is non-monotonic for the final EoR stages, when multiple bubble overlap starts being significant. Applying this non-monotonic calibration curve to the model that includes recombinations can introduce unphysical features in the EoR history.  To minimize this effect, we force the calibration curve to be monotonic, linearly extrapolating below $1-Q_{\rm \ion{H}{II}} <0.3$.  While there is still a slight knee in some EoR histories, we find this implementation is robust to a wide range of our astrophysical parameter space.}
On the other hand, when recombinations are not included, the corrected (thick red dashed line in [3] of Fig.~\ref{fig:schematic}) and analytic EoR histories (thin violet solid line) are almost identical, by construction.

\begin{figure}
    \includegraphics[width=8.5cm]{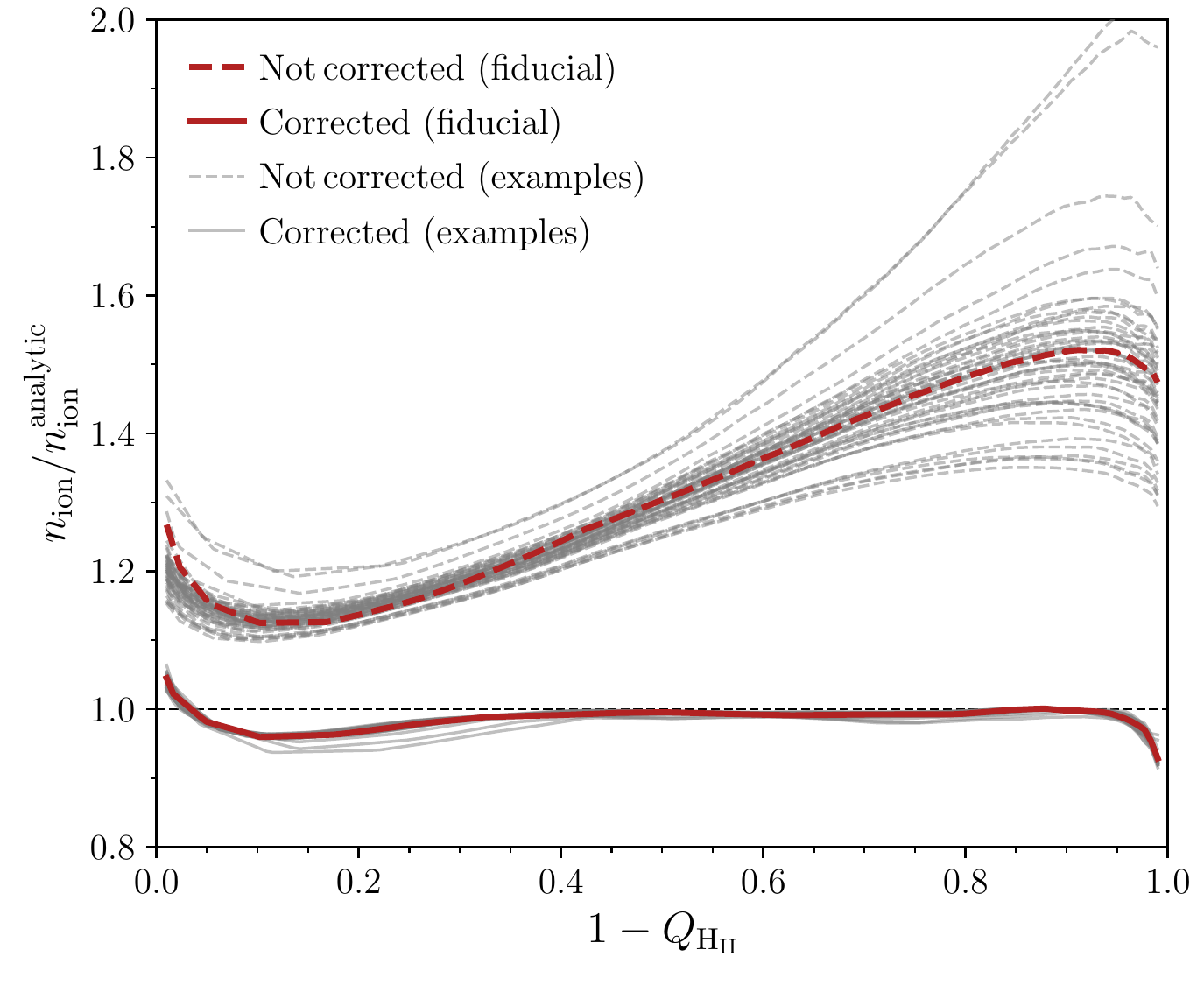}
    \caption{
    The ratio of the cumulative ionization emissivity from {\cmfast} to the cumulative ionization emissivity from the analytic expression as a function of the mass-weighted ionized filling factor ($Q_{\rm \ion{H}{II}}$). Thick dashed (red) and thick solid (red) lines represent the ratio before and after the correction for our fiducial model, respectively. Thin dashed (grey) and solid (grey) lines represent 50 different parameter combinations. We note that most of thin dashed (grey) lines are overlapping with the thick solid (red) line.
    }
\label{fig:ratio}
\end{figure}

Thus far, we have shown the impact of photon conservation on the evolution of the EoR history, $Q_{\rm \ion{H}{II}}(z)$.  Because our calibration is not perfect, the resulting misestimate of $Q_{\rm \ion{H}{II}}$ post-calibration will still result in a misestimate of the ionizing photon production rate of galaxies. In Fig. \ref{fig:ratio} we quantify this error in the limit of no recombinations from equation (\ref{eq:HII_filling_factor}), showing the ratio of $n_{\rm ion}$ inferred from the EoR history produced by {\cmfast} to the true value $n_{\rm ion}^{\rm analytic}$.
If the calibration were perfect, this ratio would be unity. As expected, the ratio deviates from unity without photon conservation (dashed lines). On the other hand, with photon conservation (solid lines), the ratio is close to unity. Over the middle stages of the EoR (which are the easiest to detect and thus provide the most constraining power; e.g. \citealp{Greig&Mesinger2017}) the ratio is indeed unity, with errors of up to a few percent at the early and late stages. This is a significant improvement compared to the the tens -- hundred percent error pre-calibration, as well as the current factor of $\sim$ 10 uncertainty on this quantity (e.g. \citealp{Bouwens2015}).


Our new recipe to approximately correct ionizing photon non-conservation increases the computing time by only $\sim 20$ per cent for a full run, including recombinations and spin temperatures. This modest additional overhead means that it can be applied {\it on-the-fly} for the Bayesian inference. And indeed it was already included in the analysis of recent 21-cm upper limits \citep{Greig2021LOFAR, Greig2021MWA, HERA2021}.

%
%
\section{Setting up Bayesian inference} 
\label{sec:mock_obs}

In order to quantify the impact of ignoring ionizing photon non-conservation, we compare the results of Bayesian parameter inference with and without applying the correction discussed in the previous section. Here we summarize the model and observations we will use in this inference.

\begin{figure*}
    \includegraphics[width=18cm]{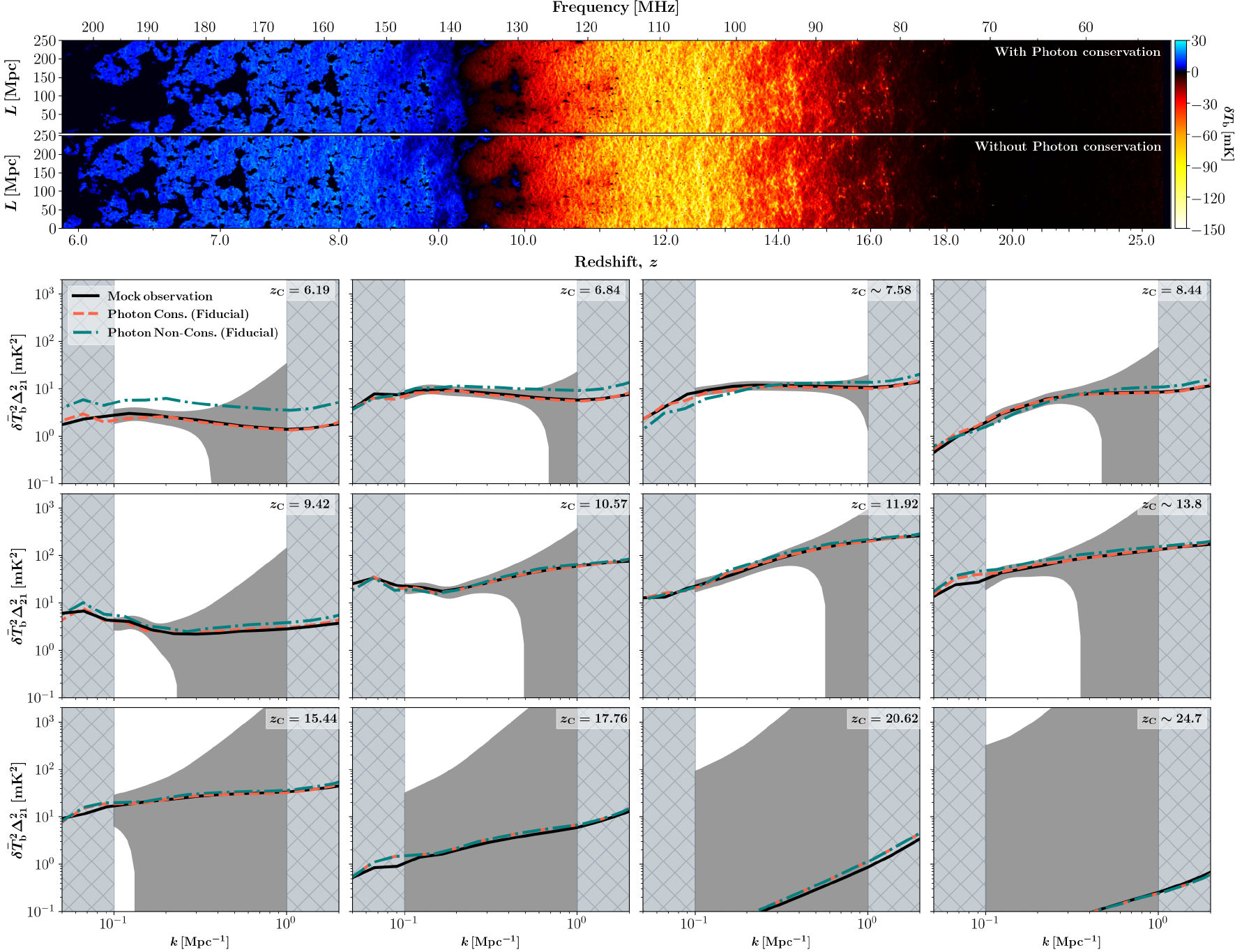}
    \caption{ Top panels: 2D light-cone slices with and without photon conservation for comparison. Both light-cones are generated from a box size of $250^3\,{\rm Mpc^3}$ using the fiducial parameter set. Bottom panels: the mock 21-cm PS (black solid) calculated from twelve independent `chunks' with $1\sigma$ noise assuming a HERA 1000h observation (shaded grey region). In each panel, hatched region outside of $k=0.1 {\rm \,Mpc^{-1}}$ and $k=1\,{\rm Mpc^{-1}}$ represent excluded regions in our likelihood calculation. Dashed (red) and dot-dashed (green) lines represent PS generated from the box size of $250^3\,{\rm Mpc^3}$ with and without photon conservation, respectively.  Neglecting ionizing photon conservation delays the EoR and the corresponding 21-cm PS evolution; as we see in the following section, this effect is mimicked by a shift in the $\alpha_{\rm esc}$ parameter.}
\label{fig:mock_obs}
\end{figure*}
\subsection{The galaxy model}\label{sec:galaxy_model}
We use the galaxy model of \cite{Park2019}, which assumes flexible, empirical scaling relations connecting  high-redshift galaxy properties to host dark matter halo masses (see also, e.g. \citealt{Kuhlen2012,Dayal2014,Behroozi2015,Mitra2015,Mutch2016,Sun&Furlanetto2016,Yue2016,Ocvirk2016,Salcido2019,Hutter2021}). This model can reproduce all current high-$z$ observables, including the rest-frame UV luminosity functions \citep[e.g.][]{Bouwens2016_LF6,Bouwens2015_LF4-10,Oesch2018}, Lyman alpha forest statistics \citep[e.g.][]{Bosman2018,Bosman2021}, and various probes of the EoR history \citep[e.g.][]{McGreer2015,Planck2016,Pagano2020}. Furthermore, the galaxy scaling relations are consistent with results from hydrodynamic simulations \citep[e.g.,][]{Xu2016, Park2020, Ma2020} and semi-analytical galaxy formation models \citep[e.g.,][]{Sun&Furlanetto2016, Mutch2016,Tacchella2018,Behroozi2019,Yung2019}. We briefly summarize the model here and its free parameters; interested readers can find more details in \citet{Park2019}.

We are only concerned with the faint end of the high-$z$ galaxy luminosity functions as these galaxies dominate the photon budget \citep[e.g.][]{Qin2021_Lya}. We thus characterize the specific stellar mass with a single power-law:
\begin{equation}\label{eq:F_STAR}
\frac{M_{\ast}}{M_{\rm h}} = f_{\ast,10} \left( \frac{M_{\rm h}} {10^{10}{\rm M}_{\sun}} \right)^{\alpha_{\ast}} \left(\frac{\Omega_{\rm b}}{\Omega_{\rm m}}\right),
\end{equation}
where $(\Omega_{\rm b}/\Omega_{\rm m})$ represents the mean baryon fraction and  $f_{\rm \ast}(M_{\rm h})=f_{\rm \ast,10}(M_{\rm h}/10^{10} {\rm M}_\odot)^{\alpha_{\ast}}$ is the fraction of the galactic gas in stars. Here, $f_{\rm \ast,10}$ and $\alpha_{\ast}$ are free parameters and $f_{\rm \ast}(M_{\rm h})$ is allowed to vary between 0 and 1. Consistent with current observations \citep[e.g.,][]{Oesch2018,Harikane2018,Bouwens2021}, the star formation efficiency is independent of redshift. The star formation rate (SFR) is characterized by a time-scale that scales with the halo dynamical time (or analogously with the Hubble time, $H(z)^{-1}$): 
\begin{equation}\label{eq:SFR}
\dot{M_{\ast}}(M_{\rm h},z) =  \frac{M_{\ast}}{t_\ast H(z)^{-1}}, 
\end{equation}
where $t_{\ast}$ is a free parameter. We  account for inefficient cooling and/or feedback processes in low-mass halos \citep[e.g][]{Shapiro1994,Giroux1994,Hui1997,Barkana&Loeb2001,Springel&Hernquist2003,Okamoto2008,Mesinger2008,Sobacchi2013a,Sobacchi2013b} by including a duty cycle (i.e. fraction of halos hosting star forming galaxies; c.f. eq. 3): $f_{\rm duty}(M_{\rm h}) = \exp\left( - M_{\rm turn}/M_{\rm h}\right)$.

We also characterize the UV ionizing escape fraction, $f_{\rm esc}$, with a power-law:
\begin{equation}\label{eq:F_ESC}
f_{\rm esc}(M_{\rm h}) = f_{{\rm esc, 10}}\left( \frac{M_{\rm h}}{10^{10}{\rm M}_{\sun}}\right)^{\alpha_{\rm esc}},
\end{equation}
where $f_{{\rm esc, 10}}$ is the normalization and $\alpha_{\rm esc}$ is a power-law index. Such a power-law scaling is consistent with population-averaged relations found by hydrodynamic RT simulations (e.g. \citealt{Paardekooper2015,Kimm2017,Lewis2020}), although there is currently no consensus on the values of the escape fraction. $f_{\rm esc}(M_{\rm h})$ is also limited to be between 0 and 1.

We account for an inhomogeneous X-ray background by parameterizing the typical emerging X-ray spectral energy distribution of high-$z$ galaxies, using the integrated soft-band ($<2{\rm keV}$) luminosity per SFR (in units of ${\rm erg\,s^{-1}\,M_{\sun}^{-1}\,yr}$),
\begin{equation}\label{eq:soft_X-ray}
L_{\rm X<2\,keV}/{\rm SFR} = \int_{E_0}^{2\,{\rm keV}}{\rm d}E_{\rm e}\, L_{\rm X}/{\rm SFR},
\end{equation}
where $E_0$ is the X-ray energy threshold below which photons are absorbed by their host galaxies. Note that we assume the specific X-ray luminosity follows a power-law in photon energy, i.e. $L_{\rm X} \propto E^{-\alpha_{\rm X}}$ and take $\alpha_{\rm X}=1$, consistent with models of High Mass X-ray Binaries in primordial galaxies (e.g. \citealt{Fragos2013, Mineo2012, Das2017}). We then compute the ionization and heating rates by integrating the angle-averaged specific X-ray intensity (in units of ${\rm erg\,s^{-1}\,keV^{-1}\,cm^{-2}\,sr^{-1}}$) back along the lightcone in each simulation cell.  In a similar fashion, the Lyman series background is computed by integrating back along the lightcone for each cell. For both X-rays and Lyman series we compute the corresponding attenuation accounting for hydrogen and helium in a multi-phase IGM \citep{21cmfast}. It is important to highlight that these radiation fields are not computed via excursion set.  Excursion set is only applicable for bi-modal fields driven by short mean free path photons, like the patchy EoR.  Photon conservation for X-ray and Lyman series photons is implicitly insured by integration along the lightcone.

In summary, our galaxy model has eight free parameters:

(1) $f_{\rm \ast,10}$: the normalization of the star formation efficiency, evaluated for a halo mass of $10^{10}M_{\rm \odot}$. We allow the parameter to vary in the range of ${\rm log_{10}}(f_{\rm \ast,10}) = [-3.0, 0.0]$.

(2) $\alpha_{\ast}$: the power-law scaling of $f_{\ast}$ with halo mass. We allow the parameter to vary in the range of $\alpha_{\ast} = [-0.5, 1.0]$.

(3) $f_{\rm esc,10}$: the normalization of ionizing escape fraction, evaluated for the halo mass of $10^{10}M_{\rm \odot}$. We allow the parameter to vary in the range of ${\rm log_{10}}(f_{\rm esc,10}) = [-3.0, 0.0]$.

(4) $\alpha_{\rm esc}$:  the power-law scaling of $f_{\rm esc}$ with halo mass. We allow the parameter to vary in the range of $\alpha_{\rm esc} = [-1.0, 0.5]$.

(5) $M_{\rm turn}$: the halo mass scale below which the abundance of star-forming galaxies is exponentially suppressed. We allow the parameter to vary in the range of ${\rm log_{10}}(M_{\rm turn})$ = [8.0, 10.0].

(6) $t_{\ast}$: the star formation time-scale as a fraction of the Hubble time, $H^{-1}(z)$. We allow the parameter to vary in the range of $t_{\ast}$ = (0.0, 1.0].

(7) $L_{\rm X<2\,keV}/{\rm SFR}$: the soft-band X-ray luminosity per unit star formation. We allow the parameter to vary in the range of ${\rm log_{10}}(L_{\rm X<2\,keV}/{\rm SFR})$ = [38.0, 42.0].

(8) $E_0$: the minimum X-ray photon energy capable of escaping the galaxy. We allow the parameter to vary in the range of $E_0$ = [0.1, 1.5].

\begin{table*}
\begin{center}
\caption{ The medians and 68\% C.I. of the eight parameter astrophysical model (see \S. \ref{sec:galaxy_model}) inferred from a mock 1,000h HERA observation, computed with the fiducial parameters and a different initial seed. The likelihood also includes existing observational constraints from UV LFs, the effective optical depth to the CMB and the dark fraction of pixels in QSO spectra.}
\begin{tabular} {ccccccccc}
\\
\hline\\[-3.0mm]
               &  &  &  &  Parameters  &  &  &       \\[1mm]
               & ${\rm log_{10}}(f_{\ast,10})$ & $\alpha_{\ast}$ & ${\rm log_{10}}(f_{\rm esc,10})$ & $\alpha_{\rm esc}$ & ${\rm log_{10}}(M_{\rm turn})$ & $t_{\ast}$ & ${\rm log_{10}}\left(\frac{L_{{\rm X}<2{\rm keV}}}{\rm SFR}\right)$ & $E_0$   \\[1mm]
               &  &  &  &  & $[{\rm M_{\sun}}]$  &   & $[{\rm erg\,s^{-1}\,M_{\sun}^{-1}\,yr}]$ &  $[{\rm keV}]$   \\[1.5mm] \hline \\[-2.5mm]
  Fiducial values  & $-1.30$ & $0.50$ & $-1.10$ & $-0.40$ & $8.70$  & $0.5$ & $40.50$ &  $0.50$   \\[1.5mm] \hline\\[-2.5mm]              
 Photon Conservation     & $-1.23^{+0.16}_{-0.20}$ &  $0.54^{+0.06}_{-0.07}$ &  $-1.17^{+0.21}_{-0.16}$  &  $-0.44^{+0.06}_{-0.06}$   &   $8.74^{+0.17}_{-0.17}$   &  $0.55^{+0.21}_{-0.19}$   &    $40.50^{+0.03}_{-0.03}$  &   $0.48^{+0.02}_{-0.02}$\\[1.5mm]
 Without photon conservation  &  $-1.22^{+0.15}_{-0.18}$ &  $0.51^{+0.06}_{-0.06}$  &  $-1.10^{+0.19}_{-0.17}$  &  $-0.62^{+0.08}_{-0.09}$  &   $8.87^{+0.13}_{-0.15}$   &  $0.53^{+0.21}_{-0.17}$   &   $40.52^{+0.03}_{-0.03}$   &   $0.48^{+0.01}_{-0.01}$\\[1.5mm]
  \hline
  
\end{tabular}
\label{Table:recovered_parameters}
\end{center}
\end{table*}

\subsection{The data}\label{sec:The_data}

To compute the likelihood, we use several existing observations. These include: (i) the electron scattering optical depth to the CMB, $\tau_{\rm e} = 0.059 \pm 0.006 (1\sigma)$ \citep{Pagano2020}, (ii) the dark fraction of pixels in QSO spectra, $x_{\ion{H}{I}} < 0.06 + 0.05 (1\sigma)$ at $z=5.9$ \citep{McGreer2015} and current rest-frame UV LFs at $z\sim 6$, 7, 8 and 10 \citep{Bouwens2016_LF6,Bouwens2015_LF4-10,Oesch2018}. We  assume Gaussian or truncated Gaussians for the likelihood shapes, depending on whether the observations are upper limits or not. For more details on these implementations see \citet{Park2019}.

In addition to these existing observations, we also make a mock detection of the 21-cm PS. Once detected, the 21-cm signal will provide much more constraining power than current observations. This combination of data will give us the tightest posterior foreseeable in the near future, allowing us to more easily identify the biases introduced by neglecting photon non-conservation.

We compute the mock cosmic 21-cm signal using the latest version \citep{21cmfast_v3} of {\cmfast} \citep{Mesinger2007,21cmfast}. The simulation box is 500 Mpc on a side with a $256^3$ grid, downsampled from $1024^3$ initial conditions\footnote{For computational efficiency, we generate the forward-modelled simulations in a volume that is smaller by a factor of $2^3$, but keeping the same resolution of $\sim 2 {\rm Mpc}$. We also use a different initial cosmic seed. See \cite{21CMMC} for more details.}. We use the fiducial parameters listed in Table~\ref{Table:recovered_parameters}. These are consistent with current observational constraints, as we show in the following section. Note that when creating the mock observation we include our correction of photon non-conservation.

\begin{figure*}
    \includegraphics[width=18cm]{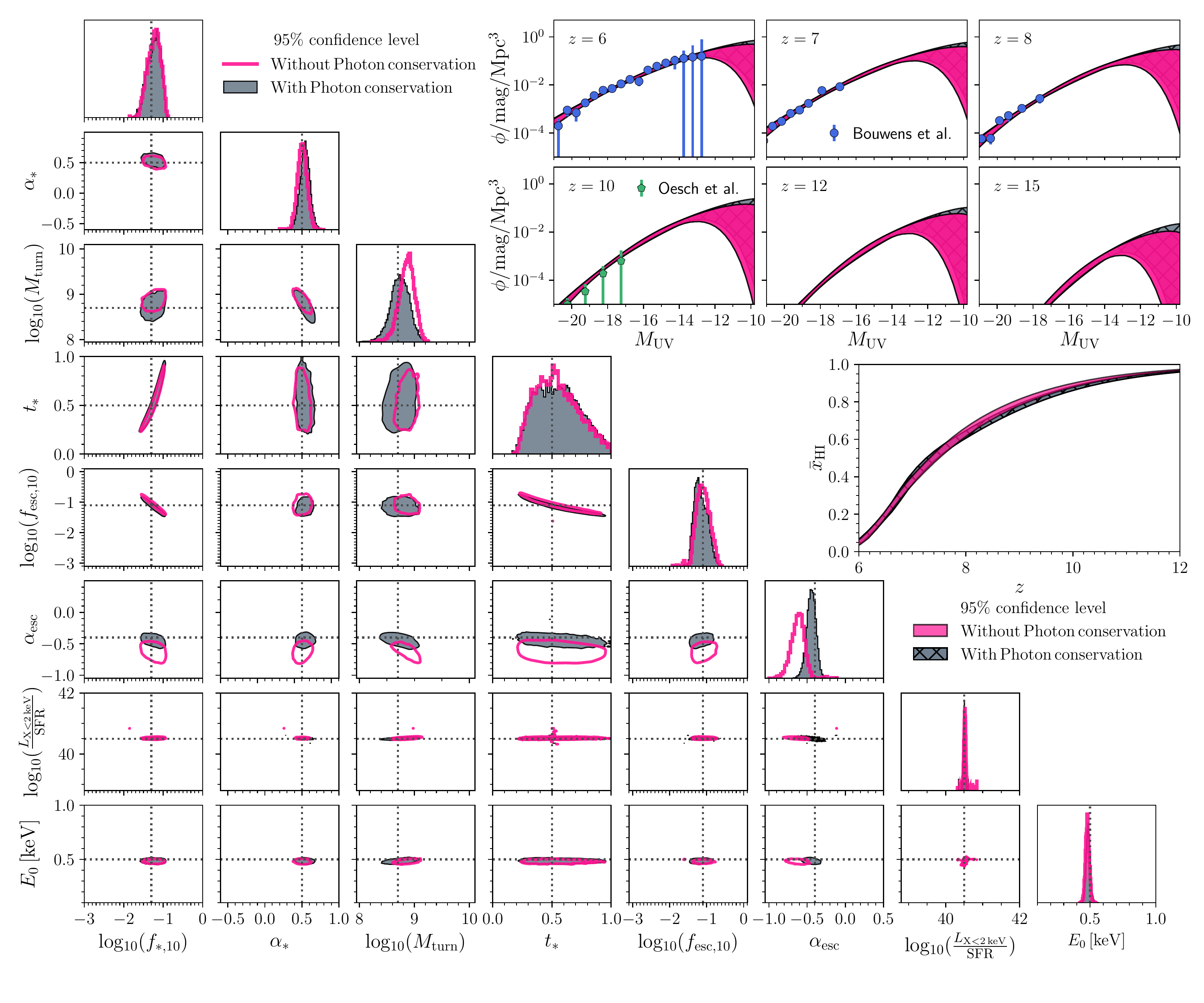}
    \caption{Marginalised posteriors with and without using our photon conservation correction (see the legend).  The mock 21-cm signal was generated with photon conservation, using the fiducial parameters denoted by the dotted lines in the corner plot. The bottom left shows the 1D and 2D parameter constraints, while in the top right we show the UV LFs and the EoR history.  The shaded regions represent 95 percent credible intervals C.I.}
\label{fig:corner_plot}
\end{figure*}

We use the spherically averaged 21-cm power spectra (PS) to characterize the 21-cm lightcone. To calculate these PS we split the simulated light-cone of our mock signal along the redshift/frequency axis in equal comoving volumes equivalent to the sampled light-cones within {\cmmc} following \cite{Greig2018}. This results in twelve independent `chunks' for which we compute the 3D PS. We consider a mock observation with HERA \citep{Beardsley2015} consisting of 331 14m dishes. To construct the thermal noise and cosmic variance on our mock observation of the PS we use \textsc{21cmsense} \citep{Pober2013,Pober2014}. We assume a 1,000 hour observation and adopt the moderate foreground removal strategy from \cite{Pober2014}. This consists of removing contaminated Fourier modes from within the foreground wedge. We also include a 20\% "model uncertainty", crudely accounting for inaccuracies in our numerical implementation \citep{Zahn2011,Ghara2015,Hutter2018}.

In Fig.~\ref{fig:mock_obs}, we show the mock 21-cm observation. The bottom panels correspond to the PS (black solid line) calculated from the twelve independent `chunks' with $1\sigma$ noise assuming a 1,000 hour observation with HERA (shaded grey region). In each panel, the hatched regions outside of $k=0.1 - 1 {\rm \,Mpc^{-1}}$ represent excluded regions in our likelihood calculation to avoid foreground contamination and shot noise, respectively. We also show 2D slices extracted from the 3D light-cone (top) and PS (bottom) generated from the smaller $250^3\,{\rm Mpc^3}$ box sampled within {\cmmc} with and without photon conservation, {\it computed for the same galaxy parameters}. As expected the differences between two models become larger as reionization proceeds. As we saw in the previous section, correcting for the lost ionizing photons shifts the EoR to earlier times, accelerating its evolution.  This shift in timing means that {\it at a fixed redshift}, the un-calibrated model has a smaller $Q_{\rm \ion{H}{II}}$ and smaller \ion{H}{II} regions, as well as a slower/delayed PS evolution.

It is important to note that the comparison in Fig.~\ref{fig:mock_obs} is done for a fixed set of galaxy parameters (the fiducial ones in Table 1). As we will see in the next section, the impact of non-conservation can be compensated by varying the galaxy parameters. Therefore, although one can recover the observations even ignoring photon non-conservation, the corresponding inferred galaxy parameters will be biased (see also \citealt{Paranjape2014, Hutter2018}).

We also note that \cite{Choudhury2018} found that large-scale 21cm power is resolution dependent for excursion set approaches. In practice, any numerical simulation is resolution dependent, because the sub grid recipies themselves depend on the scales that are resolved (recombinations, non-linear densities). Here we briefly compare the 21cm power spectra from our fiducial model (computed on a $128^3$ grid) to a lower resolution simulation (computed on a $64^3$ grid). We find that, regardless of whether or not photon conservation is used, the difference of large-scale 21cm power between the two resolution simulations is at the level of $\sim$ few percent: well within our assumed 20\% modeling uncertainty. This is also considerably smaller than the expected observational uncertainty, and so would have a negligible impact on inference.

%
%
\section{Parameter bias due to ionizing photon non-conservation}\label{sec:results}

Using our observational data sets, including the mock 21-cm PS, we perform two Bayesian inferences: with and without accounting for photon non-conservation. We use the public MCMC sampler of 3D EoR simulations {\cmmc}\footnote{https://github.com/21cmfast/21CMMC} \citep{21CMMC,Greig2017,Greig2018}. At each MCMC step, {\cmmc} calls {\cmfast} to generate the lightcone of the cosmic 21-cm signal, the UV LFs, as well as other observables.

Our two posteriors are shown in Fig.~\ref{fig:corner_plot}, with gray/red corresponding to with/without photon conservation. The 1D and 2D marginalized PDFs are shown in the lower left, while the UV LFs and inferred EoR histories are shown in the top right. Shaded regions correspond to 95\% credible intervals (C.I.). The recovered median and 68\% C.I. for all parameters are also listed in Table \ref{Table:recovered_parameters}.

In general, we find the biases in the inferred galaxy properties to be modest. The largest impact is in the recovery of $\alpha_{\rm esc}$, which is underestimated by $\sim 0.2$ (corresponding to $\sim 2.4\sigma$). This is understandable since in our parameterization, $\alpha_{\rm esc}$ is most responsible for the speed of reionization, given that the star formation rate parameters are constrained by the UV LFs (c.f. \citealt{Park2019}). Thus a decrease in $\alpha_{\rm esc}$ can mimic the effect of ionizing photon conservation.  Indeed we see that the recovered EoR histories are very similar in the two models.

We also find the recovered median value of ${\rm log_{10}}(M_{\rm turn})$ from the model without photon conservation to be larger than that of the model with photon conservation by 1.5 percent (35 percent on a linear scale). This parameter also can impact the speed of reionization, though current UV LFs provide an upper limit.  However, this bias is very small and the recovered 95\% limits are fully consistent between the two posteriors.

%
%
\section{Conclusions}\label{sec:conclusion}

While excursion set approaches to generate the 3D ionization structure are well tested and efficient, the ionizing photon number is not conserved. Here, we introduce a new recipe for approximately correcting the photon non-conservation that can be applied on-the-fly within {\cmfast}. Our approach preserves the well-tested reionization topology of excursion set methods, simply shifting the ionizing field in redshift to compensate for photon non-conservation. We find that this calibration is quite insensitive to astrophysical parameters, resulting in shifts of order $\Delta z(Q_{\rm \ion{H}{II}})\sim 0.1$.  It is also fast, facilitating its application to high-dimensional Bayesian inference.

Using current observations and a mock 21-cm detection, we quantify the bias in astrophysical parameter recovery if photon non-conservation is ignored. For our galaxy model and fiducial parameter choice, we find this bias to be very small for most parameters. The major exception is for the power-law scaling of the ionizing escape fraction with halo mass, $\alpha_{\rm esc}$.  Without correcting for photon conservation, the recovered values of the power law index are biased low by 0.2, corresponding to $2.4 \sigma$.  However, the EoR histories are largely consistent in the two models, demonstrating that a bias in $\alpha_{\rm esc}$ is sufficient to compensate for ionizing photon non-conservation of excursion set models.

Our approximate correction for ionizing photon non-conservation has been publicly available since v3.0 of {\cmfast} (through the flag \verb!`PHOTON_CONS' -> True!).

\section*{Acknowledgements}
JP was supported by a KIAS Individual Grant (PG078702) at Korea Institute for Advanced Study. Parts of this research were supported by the Australian Research Council Centre of Excellence for All Sky Astrophysics in 3 Dimensions (ASTRO 3D), through project number CE170100013.

\section*{DATA AVAILABILITY}
The data underlying this article will be shared on reasonable request to the corresponding author.



\bibliographystyle{mnras}
\bibliography{ref.bib} 







\bsp	
\label{lastpage}
\end{document}